\documentclass{article}
\usepackage{spconf,amsmath,graphicx, url}
\usepackage[dvipsnames]{xcolor}
\usepackage{multirow}
\usepackage{tabularx}
\usepackage[shortcuts]{extdash}
\usepackage{enumitem}
\setlist{nosep, leftmargin=14pt}

\usepackage{mwe} 

\title{Deep Learning Based Segmentation of Blood Vessels from H\&E Stained Oesophageal Adenocarcinoma Whole-Slide Images}
\name{Jiaqi Lv$^{\star}$ \qquad Stefan S Antonowicz$^{\dagger}$ \qquad Shan E Ahmed Raza$^{\star}$}
\address{$^{\star}$ Tissue Image Analytics Centre, Department of Computer Science, University of Warwick. \\
    $^{\dagger}$ Department of Surgery \& Cancer, Faculty of Medicine, Imperial College London.}
\begin{document}
%
\maketitle
\begin{abstract}
Blood vessels (BVs) play a critical role in the Tumor Micro-Environment (TME), potentially influencing cancer progression and treatment response. However, manually quantifying BVs in Hematoxylin and Eosin (H\&E) stained images is challenging and labor-intensive due to their heterogeneous appearances. We propose a novel approach of constructing guiding maps to improve the performance of state-of-the-art segmentation models for BV segmentation, the guiding maps encourage the models to learn representative features of BVs. This is particularly beneficial for computational pathology, where labeled training data is often limited and large models are prone to overfitting. The quantitative and qualitative results demonstrate the efficacy of our approach in improving segmentation accuracy. In future, we plan to validate this method to segment BVs across various tissue types and investigate the role of cellular structures in relation to BVs in the TME.
\end{abstract}
\section{Introduction}
\label{sec:intro}
Blood vessels (BVs) in the Tumor Micro-Environment (TME) play a significant role in cancer progression and treatment response. One of the hallmarks of cancer is inducing angiogenesis, a process through which tumors stimulate the formation of new blood vessels to sustain their growth. These vessels provide essential nutrients and oxygen while removing waste products from the TME \cite{Hanahan2011}. H\&E staining is a standard method used by pathologists for tissue examination. However, due to the heterogeneity in the size, shape, and distribution of BVs, manual quantification of BVs in H\&E-stained images is a challenging, time-consuming, and subjective task. This challenge is especially pronounced when dealing with multi-gigapixel whole-slide images (WSIs). As a result, pathologists do not routinely quantify BVs, which limits our understanding of their role in the TME.

Recent efforts in automating the segmentation of BVs in histology images using deep learning have shown promise, though they have some limitations. For instance, Glanzer et al. \cite{Glanzer2023} trained a U-Net model \cite{Ronneberger2015} to segment BVs in pancreatic tumor images. To tackle the challenge of BV heterogeneity, their method requires manual annotation of six types of structures: vessel lumen, vessel wall, normal tissue, destroyed tissue, background and debris, which is very labor-intensive. Fraz et al. \cite{warwick129881} introduced FABnet, a model designed to simultaneously segment micro-vessels and nerves in oral squamous cell carcinoma images, their method demonstrated state-of-the-art (SOTA) performance. However, one drawback of their method is the rigid pipeline and network architecture that makes it difficult to implement and extend to other tasks.   
Yi et al. \cite{yi2018microvessel} developed a fully convolutional neural network (FCN) for BV segmentation, applying it to images from 88 lung adenocarcinoma patients, they found an association between higher micro-vessel density and improved survival, but the p-value was only marginally significant. Hamidinekoo et al. \cite{pmlr-v156-hamidinekoo21a} proposed a Generative Adversarial Network (GAN) that generates synthetic ERG-stained images from H\&E input, highlighting blood vessel structures, however, this method requires ERG stained images registered with H\&E images for training, which can often be difficult to obtain. In summary, there is an emerging interest and clinical value in using deep learning to automate the task of BVs quantification in cancer biology studies, although no single superior method has been identified.

In this paper, we propose a novel method for segmenting BVs from H\&E-stained WSIs. We show that our approach improves the performance of segmentation models both qualitatively and quantitatively, by introducing guiding maps that help models better capture representative features of BVs. 

\section{Method}
\subsection{The overall framework}
\begin{figure}[t]

\begin{minipage}[b]{1.0\linewidth}
  \centering
  \centerline{\includegraphics[width=1.0\linewidth]{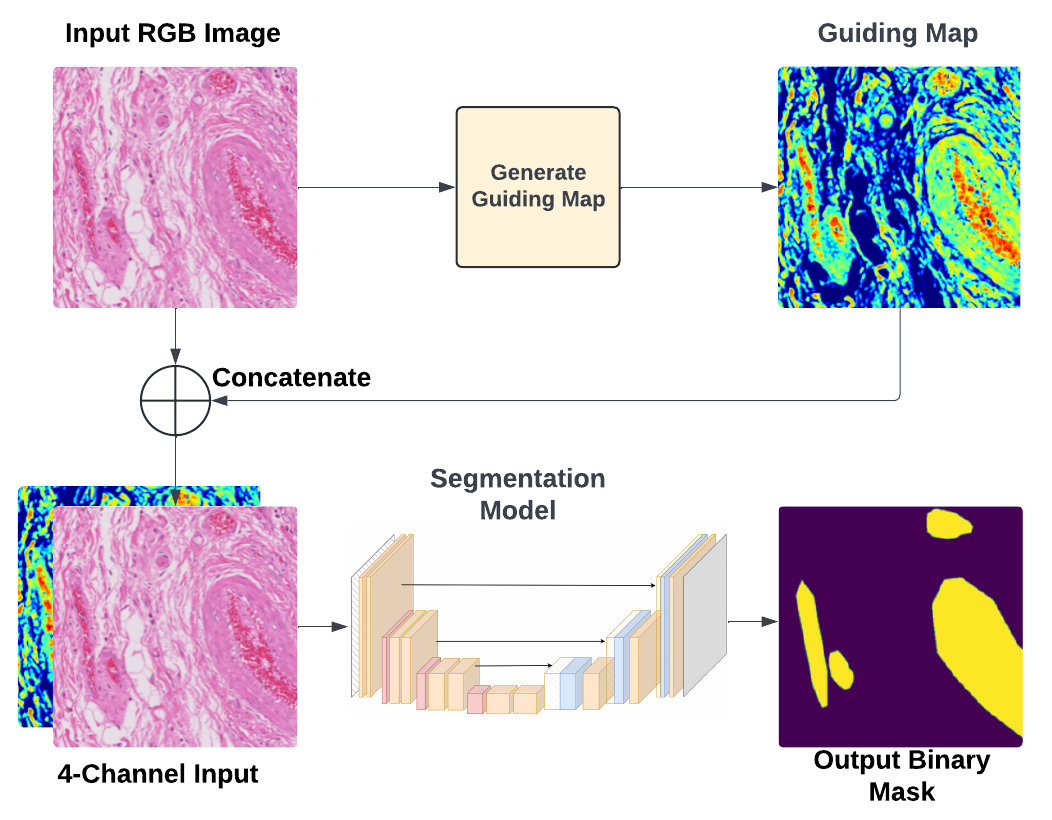}}
\end{minipage}
\caption{An overview of our proposed segmentation pipeline demonstrating how a guiding map is integrated into the input data to improve segmentation performance.}
\label{fig:pipeline}
\end{figure}

We propose a novel image processing algorithm for blood vessels (BVs) segmentation in H\&E-stained images. BVs are typically identified by the presence of red blood cells enclosed within a thin endothelial membrane. However, the presence of red blood cells inside the membrane is inconsistent, as blood may be drained during tissue processing or spilled onto other parts of the tissue. This variability poses challenges for accurate BV detection and segmentation, especially in tumor areas where the shape of BVs can become very irregular. Our algorithm tackles these challenges by leveraging the shape and color characteristics of BVs, generating a high-contrast guiding map from the RGB image. This map is concatenated to the original RGB channels, forming a 4-channel input that highlights relevant features for segmentation, An overview is shown in Fig.\ref{fig:pipeline}. 

\begin{figure*}[t]

\begin{minipage}[b]{1.0\linewidth}
  \centering
  \centerline{\includegraphics[width=0.9\linewidth]{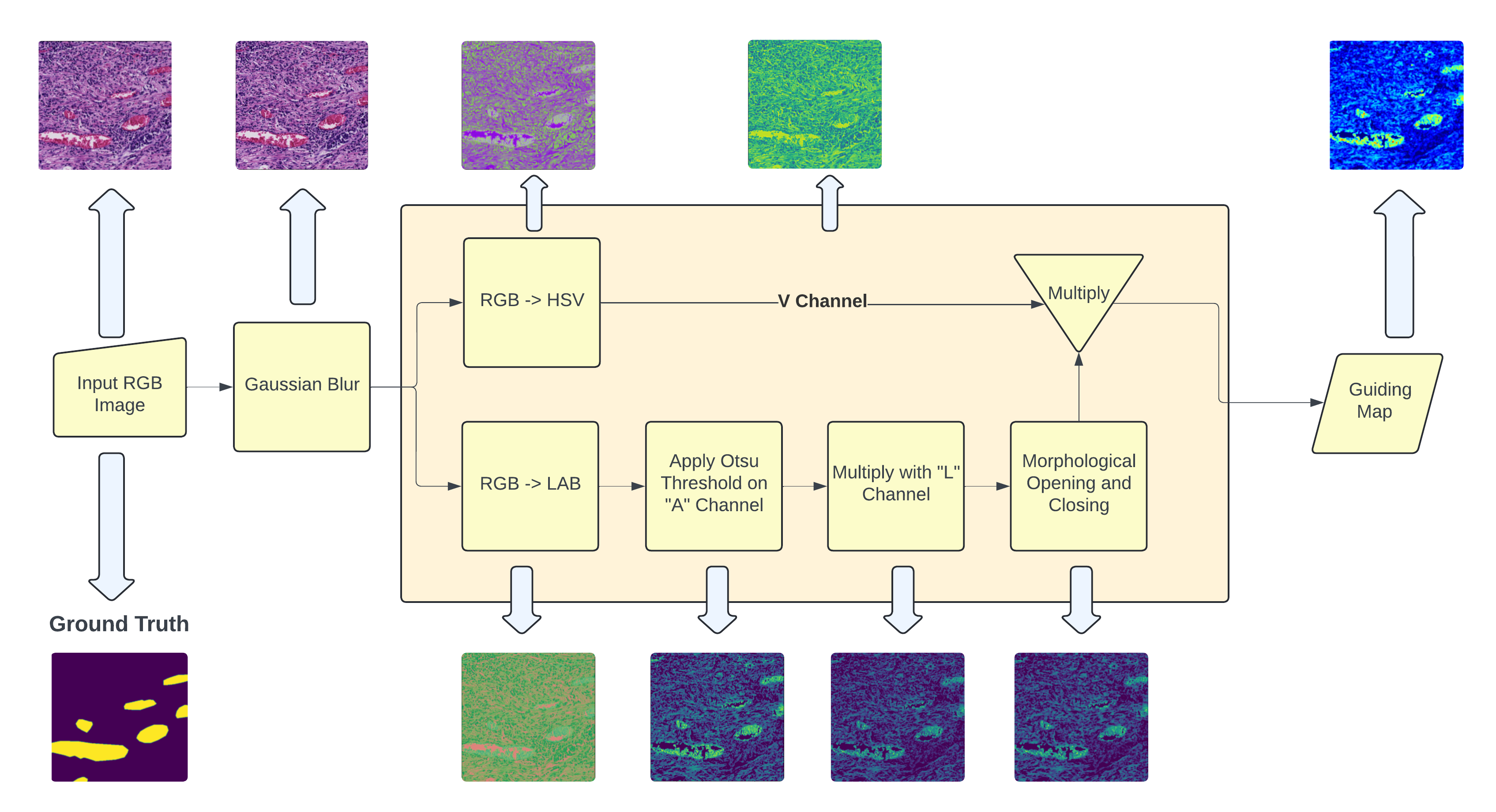}}
\end{minipage}
\caption{Algorithm to generate a guiding map from an RGB image.}
\label{fig:guiding-map-algo}
\end{figure*}

\subsection{Guiding map algorithm}

\begin{figure}[t]

\begin{minipage}[b]{1.0\linewidth}
  \centering
\centerline{\includegraphics[width=1.0\linewidth]{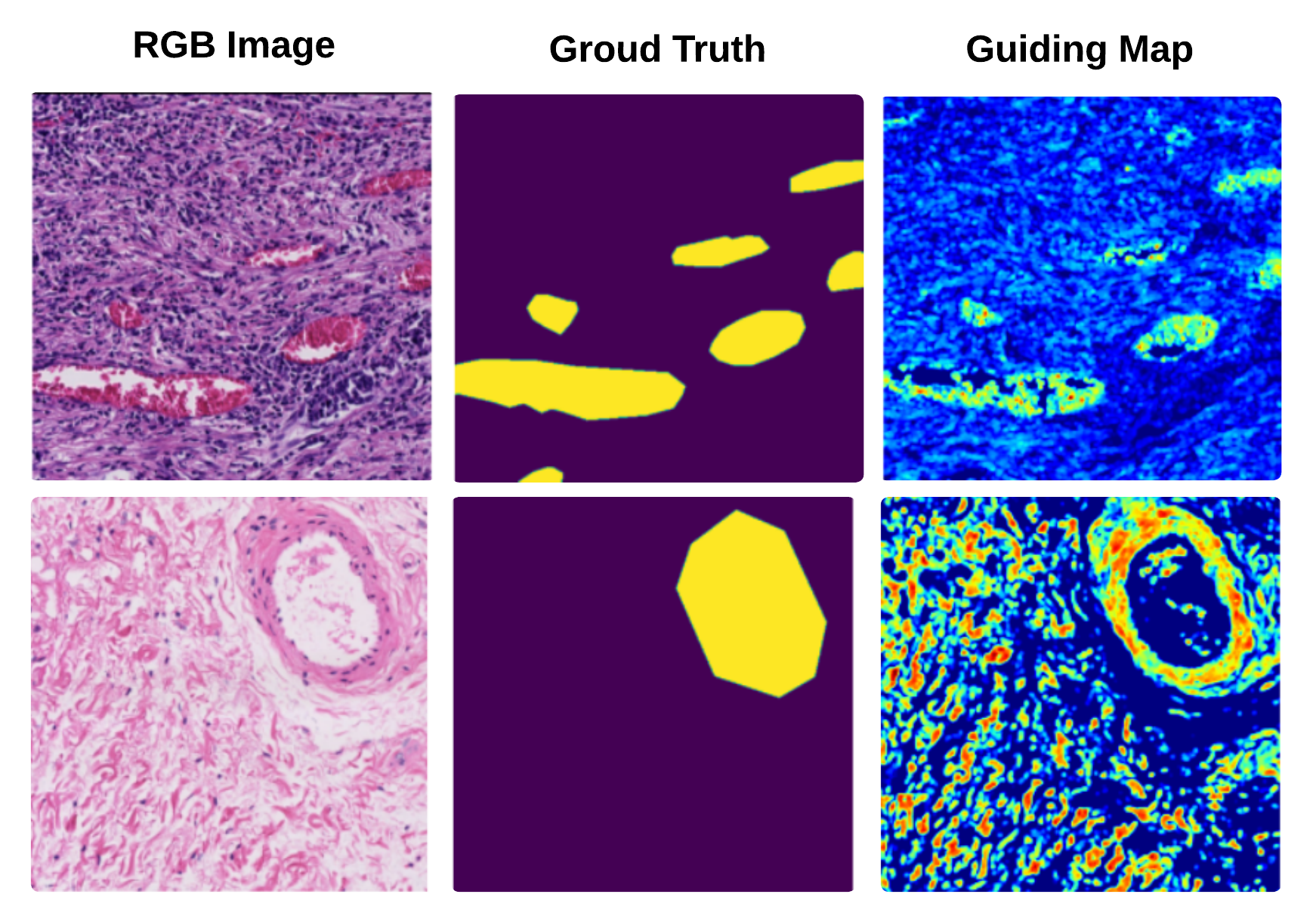}}
\end{minipage}
\caption{Examples of guiding maps. Left: Input RGB images; Middle: Ground truth binary masks of BVs; Right: Guiding maps.}
\label{fig:guiding-map-example}
\end{figure}

\begin{enumerate}
    \item \textbf{Color Space Conversions:} Given an input RGB image $x_{rgb}$ of shape $(H \times W \times 3)$, we apply Gaussian Blur to smooth the image (k=3, $\sigma$=0), then we convert it into two additional color spaces: HSV and LAB. The HSV color space represents Hue, Saturation, and Value, while the LAB color space represents Lightness (L), and the opponent color channels A (green-red) and B (blue-yellow). This conversion produces two new images: $x_{hsv}$ and $x_{lab}$.
    \item \textbf{Thresholding with the A Channel:} From $x_{lab}$, we extract the A channel, $x_{a}$ of shape $(H \times W \times 1)$. Using Otsu’s thresholding method, we obtain the optimal threshold $t$ that distinguishes red-colored pixels, then $t$ is subtracted from $x_{a}$ to obtain $x_{a'}$, any negative values are set to zero. This step generates a heat-map of red pixels. 
    \item \textbf{Luminosity Scaling}: $x_{a'}$ is multiplied with the L channel from $x_{lab}$ to obtain $x_{a''}$, the aim is to scale the heat-map by luminosity.
    \item \textbf{Morphological Operations:} We apply morphological opening and closing operations on $x_{a''}$ (k=3). The aim is to remove small objects and small holes.
    \item \textbf{Brightness Scaling:} We further enhance the contrast in brightness of $x_{a''}$ by multiplying it with the V channel from $x_{hsv}$, which scales brightness linearly to obtain $x_{a'''}$.
    \item \textbf{Normalization:} We apply Min-Max normalization to scale each pixel value in $x_{a'''}$ to the range of [0,1] obtaining the guiding map which outlines potential blood vessels.
\end{enumerate}
Fig. \ref{fig:guiding-map-algo} illustrates the steps involved in generating a guiding map. Fig. \ref{fig:guiding-map-example}
shows examples of the resulting guiding maps.


\section{Experiments}
\subsection{Dataset and metrics}
The dataset for this study was obtained from Imperial College London where H\&E stained WSIs of oesophageal tissue were collected. A total of 28 regions of interest (ROIs) were selected from 10 different WSIs with the aim to represent the diverse characteristics of BVs from both tumoral and normal tissue regions. Blood Vessels (BVs) within these ROIs were manually annotated by an expert. From the annotated ROIs, we extracted 752 image patches, each with a size of $512 \times 512$ pixels, at $10\times$ magnification level. We split the dataset into training, validation, and test sets such that 560 patches were used for training, and 140 patches were used for validation. Both the training and validation sets were extracted from 7 WSIs, the test set containing 52 patches were extracted from 3 different WSIs.
All experiments utilized 5-fold cross-validation. Dice Similarity Coefficient (DSC) and Intersection over Union (IoU) were used for evaluating the performance of the BVs segmentation models.

\subsection{Implementation details}
Guiding maps were generated using OpenCV, and segmentation models were trained using PyTorch, initialized with ImageNet pre-trained weights. Models were trained on NVIDIA A10 GPUs using Adam optimizer (initial learning rate: 0.003, weight decay: 0.0004), Jaccard loss function and ReduceLROnPlateau scheduler. Training was conducted with a batch size of 4 over 800 epochs.

Data augmentation was applied during training using random HSV and RGB shifts, Gaussian blur, sharpening, and image compression, as well as random adjustments to brightness, contrast, shift, scale, rotation, and flipping. In training, guiding maps were generated after data augmentation.

The training dataset had a class-imbalance problem, with fewer patches containing blood vessels compared to those without. We employed the \textit{WeightedRandomSampler} from PyTorch, which ensures that each batch contained an equal number of patches with and without blood vessels.

\subsection{Comparative experiments}
To evaluate the effectiveness of our guiding map, we performed comparative experiments using the same architecture family (i.e. U-Net). We first trained the networks using RGB input, then we trained the same networks using 4-channel input, where guiding maps were concatenated to the RGB images. In our experiments, we trained U-Net with different encoders: EfficientNet B0, B2, B6 \cite{efficientnet} and EfficientNetV2-M \cite{efficientnetv2} (we name these EfficientUnet). We incorporated Spatial and Channel Squeeze \& Excitation (SCSE) blocks \cite{SCSE} into the decoder layers of the EfficientUnet models, these models were implemented using the PyTorch library \cite{Iakubovskii:2019}. Basic U-Net \cite{Ronneberger2015}, FCN \cite{yi2018microvessel} and TransUnet (R50-ViT-B\_16) \cite{chen2021transunet} were also trained on our dataset to establish benchmarks.

\section{Results and Discussion}

\begin{figure}[t]
\begin{minipage}[b]{1\linewidth}
  \centering
  \centerline{\includegraphics[height=6cm]{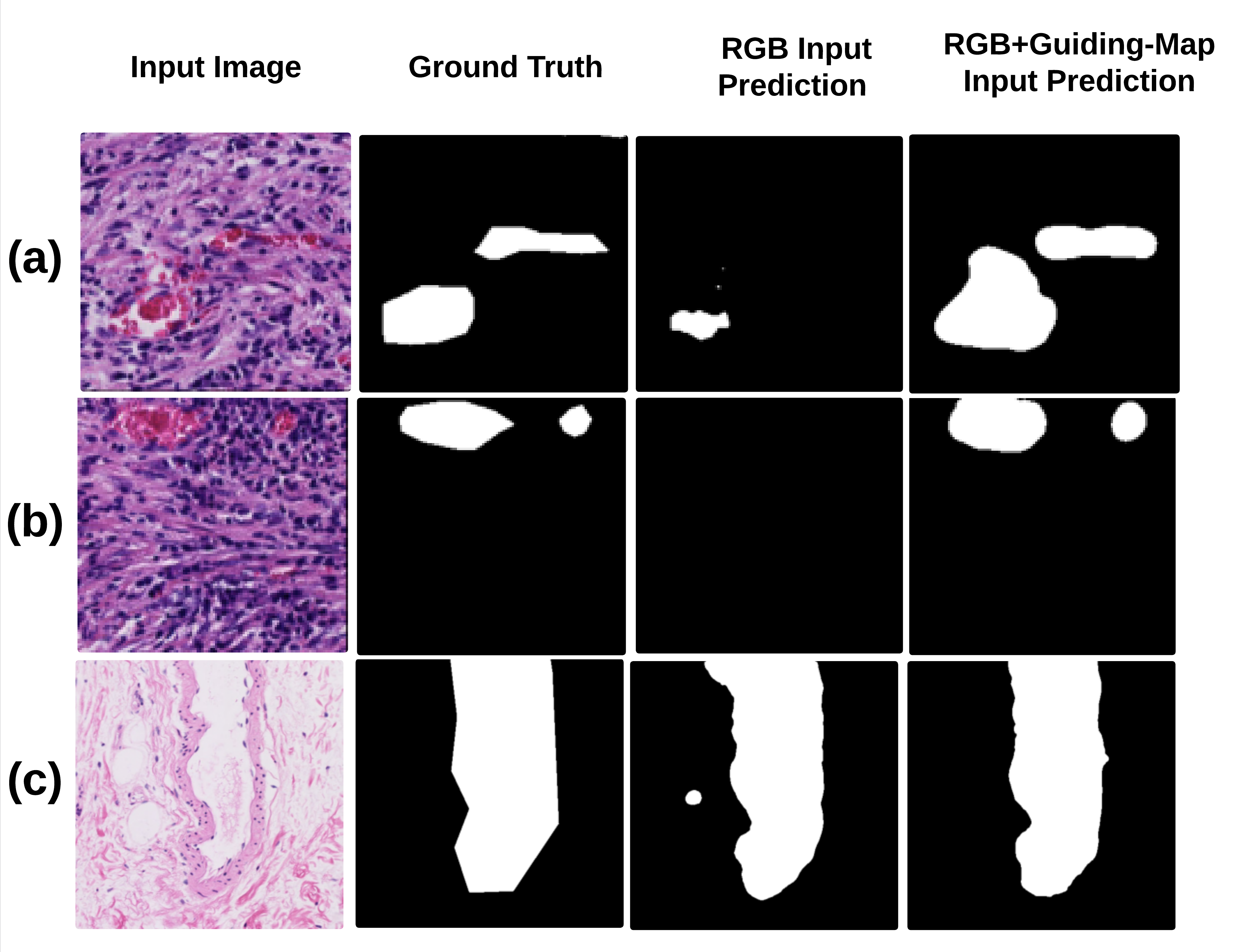}}
\end{minipage}
\caption{A comparison of segmentation quality using EfficientUnet\-/B2 model with RGB input VS with RGB+Guiding Map input. (a)(b): Tumor BVs, (c): normal BV}
\label{fig:quality-comparison}
\end{figure}

\begin{table}[t]
\centering
\small
\begin{tabular}{|c c c c|} 
 \hline
 Model & Input & DSC(\%) & IoU(\%)\\ [0.5ex] 
 \hline
 Basic Unet\cite{Glanzer2023} & \multirow{7}{3.5em}{RGB} & 77.64±2.94 & 70.90±2.55\\
 FCN\cite{yi2018microvessel} & & 75.90±1.74 & 68.76±1.82\\
 TransUnet &  & 65.09±3.84 & 58.62±3.51\\ 
 EfficientUnet-B0 & & 84.13±0.628 & 77.27±0.73\\
 EfficientUnet-B2 & & 82.06±0.724 & 75.57±0.54\\
 EfficientUnet-B6 & & 81.23±3.56 & 74.54±3.21\\
 EfficientUnet-V2-M & & 81.72±1.78 & 75.09±1.58\\
 \hline
 Basic Unet &\multirow{4}{3.5em}{RGB $+$ Guiding Map}& 81.82±2.00 $\uparrow$ & 75.34±1.88 $\uparrow$\\
 EfficientUnet-B0 &  & \textbf{85.16±1.64} $\uparrow$ & \textbf{78.39±1.71} $\uparrow$\\
 EfficientUnet-B2 & & 84.10±1.07 $\uparrow$ & 77.75±1.07 $\uparrow$\\
 EfficientUnet-B6 &  & 84.94±1.61 $\uparrow$ & 78.00±1.57 $\uparrow$\\
 EfficientUnet-V2-M &  & 84.01±1.09 $\uparrow$ & 77.06±1.63 $\uparrow$\\[1ex] 
 \hline
\end{tabular}
\caption{\textbf{Overall} segmentation performance by different methods on the test set. (FCN and TransUnet were omitted for RGB+Guiding Map because of their poor performance.)}
\label{table:1}
\end{table}

\begin{table}[t]
\small
\centering
\begin{tabular}{|c c c c|} 
 \hline
 Model & Input & DSC(\%) & IoU(\%)\\ [0.5ex] 
 \hline
 Basic Unet \cite{Glanzer2023} &\multirow{7}{3.5em}{RGB} & 69.58±9.66 & 58.20±9.31\\
 FCN \cite{yi2018microvessel} & & 79.12±3.39 & 67.60±3.92\\
 TransUnet & & 74.89 ± 4.04 & 63.73±4.28\\
 EfficientUnet-B0 & & 82.50±2.19 & 71.88±2.89\\
 EfficientUnet-B2 & & 82.37±8.35 & 73.87±7.08\\
 EfficientUnet-B6 & & 85.81±1.34 & 76.58±1.62\\
 EfficientUnet-V2-M & & 80.37±2.70  & 69.56±2.76\\
 \hline
 Basic Unet &\multirow{5}{3.5em}{RGB $+$ Guiding Map} & 79.96±6.10 $\uparrow$ & 69.14±6.31 $\uparrow$\\
 EfficientUnet-B0 &  & 83.59±3.73 $\uparrow$ & 73.88±3.50 $\uparrow$\\
EfficientUnet-B2 & & \textbf{87.37±1.25} $\uparrow$ & \textbf{78.49±1.87} $\uparrow$\\
 EfficientUnet-B6 &  & 85.61±1.25 & 75.47±1.74\\
 EfficientUnet-V2-M &  & 83.91±1.91 $\uparrow$ & 73.30±2.83 $\uparrow$\\[1ex] 
 \hline
\end{tabular}
\caption{Segmentation performance of \textbf{tumor BVs} from the test set by different methods.}
\label{table:2}
\end{table}

\begin{table}[t]
\small
\centering
\begin{tabular}{|c c c c|} 
 \hline
 Model & Input & DSC(\%) & IoU(\%)\\ [0.5ex] 
 \hline
  Basic Unet \cite{Glanzer2023} &\multirow{7}{3.5em}{RGB} & 79.56±1.54 & 73.92±1.31\\
  FCN \cite{yi2018microvessel} & & 75.13±1.67 & 69.04±1.69\\
  TransUnet & & 62.76±5.28 & 55.89±4.91\\
 EfficientUnet-B0 & & 84.52±0.45 & 78.56±0.57\\
 EfficientUnet-B2 & & 81.99±1.66 & 75.98±1.61\\
 EfficientUnet-B6 & & 80.10±3.94 & 74.06±3.99\\
 EfficientUnet-V2-M & & 82.04±1.68 & 76.40±1.60\\
 \hline
 Basic Unet &\multirow{5}{3.5em}{RGB $+$ Guiding Map} & 82.79±0.89 $\uparrow$ & 77.34±0.86 $\uparrow$\\
 EfficientUnet-B0 & & \textbf{86.09±1.98} $\uparrow$ & \textbf{80.19±1.96} $\uparrow$\\
EfficientUnet-B2 & & 83.86±1.27 $\uparrow$ & 78.21±1.23 $\uparrow$\\
 EfficientUnet-B6 &  & 85.06±1.86 $\uparrow$ & 78.82±1.71 $\uparrow$\\
 EfficientUnet-V2-M &  & 84.77±1.13 $\uparrow$ & 78.81±1.52 $\uparrow$\\[1ex] 
 \hline
\end{tabular}
\caption{Segmentation performance of \textbf{normal BVs} from the test set by different methods.}
\label{table:3}
\end{table}


The test dataset contains both normal BVs and tumor BVs. Qualitative results, are shown in Fig.\ref{fig:quality-comparison} which demonstrate that the models trained with RGB + guiding map achieve superior segmentation quality compared to those trained with RGB input only. These models produced smoother segmentation masks, with fewer irregularities and a reduction in false positive predictions while being more sensitive to tumor BVs.

The overall quantitative performance is summarized in Table \ref{table:1}. Incorporating guiding maps into the input to segmentation models resulted in an increase in overall performance by nearly \textbf{1-3\%} in DSC/IoU, with more substantial improvements observed in segmenting tumor BVs. The most significant overall improvements were observed in the EfficientUnet\-/B6 and EfficientUnet\-/V2\-/M models, while the smallest increase was observed in the EfficientUnet\-/B0 model.

Further detailed examination reveals that the use of guiding maps significantly increases the performance of EfficientUnet\-/B2 model in segmenting tumor BVs with an average increase of nearly \textbf{5\%} in DSC/IoU. This improvement was also observed across other models, except for EfficientUnet\-/B6 where the performance remains approximately the same (see Table \ref{table:2}). For segmenting normal BVs, all models demonstrated improved performance when RGB + guiding map input was used, as shown in Table \ref{table:3}.

These findings suggest that the integration of the proposed guiding map enhances the ability of deep learning models to capture representative features of both normal and tumor BVs, especially in models with higher number of parameters. This further highlights the utility of the guiding maps in reducing overfitting, particularly when training data is limited and/or contains significant variability.

\section{Conclusion and future work}
We introduced a novel image processing algorithm to enhance deep learning-based segmentation of blood vessels (BVs) in H\&E-stained oesophageal adenocarcinoma whole-slide images. By generating high-contrast guiding maps and concatenating them to RGB images, our method improves segmentation performance while requiring minimal changes to network architectures. The experimental results show significant improvements, especially in larger models, and when training data is limited. This highlights the significance of guiding maps in improving the ability of models to segment BVs.

Future work will focus on collecting a larger dataset of H\&E images registered with IHC stained images across diverse tissue types to assess the generalizability of our method and explore the relationship between BV density, spatial distribution, tumor metabolic clusters, and their correlations with prognosis and treatment response.

\section{Compliance with ethical standards}
\label{sec:ethics}
The ethical approvals were obtained from Imperial College Tissuebank REC: 22/WA/0214, Project R23042-1A.

\section{Acknowledgments}
\label{sec:acknowledgments}
SA is supported by the NIHR Imperial Biomedical Research Centre, Cancer Research UK, and the Rosetrees Trust (ref: PGL23/100123). JL is supported by the UK Engineering and Physical Sciences Research Council (EPSRC). SEAR report financial support provided by the MRC (MR/X011585/1) and the BigPicture project, funded by the European Commission.


\bibliographystyle{IEEEbib}
\bibliography{main}
\end{document}